\documentclass[twocolumn,floatfix,aps,prd,superscriptaddress,nofootinbib]{revtex4-2}

\usepackage{iftex}
\ifPDFTeX
  \usepackage[utf8]{inputenc}
  \usepackage[T1]{fontenc}
  \usepackage{lmodern}
  \input{glyphtounicode}
\else
  \usepackage{fontspec}
\fi
\usepackage{graphicx}
\usepackage{amssymb}
\usepackage{amsmath}
\usepackage{xcolor}
\usepackage{hyperref}
\usepackage{textcase}
\usepackage{placeins}
\usepackage[protrusion=true,expansion=false]{microtype}

\graphicspath{{../../../results/bondi-dipole-runaway/figures/}{figures/}}

\hypersetup{
    colorlinks=true,
    linkcolor=blue,
    filecolor=magenta,
    urlcolor=blue,
    citecolor=blue,
}
\begin{document}

\title{The Bondi dipole in full numerical relativity: a self-accelerating positive--negative mass binary}

\author{Nikita M. Shirokov}
\email{shirokov.nm@phystech.edu}
\affiliation{\textit{Moscow Institute of Physics and Technology, Moscow, Russia}}
\author{Ilya Nachevsky}
\email{ommnammashivaya@gmail.com}
\affiliation{\textit{Higher School of Economics, Moscow, Russia}}
\date{\today}

\begin{abstract}
Bondi showed in 1957 that a body of negative active gravitational mass and a positive one should accelerate together without limit while their total momentum stays zero. We test this prediction in full $3{+}1$ numerical relativity, from constraint-satisfying initial data. The matter is two complex scalar fields with identical self-interactions, one of which enters the Einstein equations with the opposite sign; its stars have positive inertial and passive masses but negative active and Arnowitt--Deser--Misner (ADM) masses. To our knowledge, they are the first asymptotically flat bodies of negative ADM mass evolved in numerical relativity. A pair of stars with equal and opposite ADM masses accelerates as a single object at constant separation, and its total momentum stays zero to within 0.5\%. The acceleration follows Newton's law of gravitation in both mass and separation, reverses when the stars are exchanged, and is absent in same-sign pairs and isolated stars. We detect no gravitational radiation. The results hold for head-on pairs released from rest, up to a few percent of the speed of light, and in the longest run the pair reaches a tenth of it with both stars intact.
\end{abstract}

\maketitle

\section{Introduction}
\label{sec:intro}

\begin{figure}[t]
\centering
\includegraphics{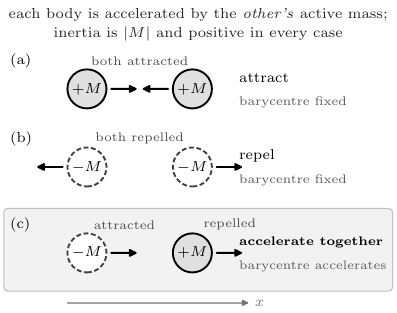}
\caption{Bondi's sign rules for two bodies released from rest, after Forward~\cite{forward1990}. Every body has positive inertial and passive mass; the active mass of the dashed bodies is negative. (a)~Two positive masses attract and (b)~two negative masses repel; in both cases the forces are opposite and the barycentre stays at rest. (c)~For a positive and a negative mass both forces point the same way, so the pair accelerates while its total momentum stays zero.}
\label{fig:schematic}
\end{figure}

Bondi~\cite{bondi1957} distinguished three kinds of mass in general relativity: the inertial mass, which resists acceleration; the passive gravitational mass, on which a gravitational field acts; and the active gravitational mass, which sources the field. He showed that a body whose active mass alone is negative is attracted by a positive mass, while the positive mass is pushed away from it. Released together, the two accelerate in the same direction without limit (Fig.~\ref{fig:schematic}), and momentum is conserved because the negative body's momentum counts with a negative sign. Bonnor and Swaminarayan~\cite{bonnor_swami1964} found exact solutions for uniformly accelerated pairs of this kind, later generalized in Ref.~\cite{bicak1983}. Forward~\cite{forward1990} showed that the self-accelerating pair conserves both momentum and energy, for equal and unequal mass magnitudes. Newtonian $N$-body simulations have since applied the same sign rules to cosmological structure formation~\cite{manfredi2018,manfredi2026}.

Nothing in the Einstein equations requires mass to be positive. Positivity of the total mass is a theorem only for matter that satisfies energy conditions~\cite{schoen_yau1979,witten1981}, and several candidate sources violate them: Casimir energies, quantum fields in curved spacetime and phantom dark energy~\cite{caldwell2002}. Traversable wormholes~\cite{morris_thorne1988} and warp-drive spacetimes~\cite{alcubierre1994} require such matter. The ghost scalar field, a field with a wrong-sign kinetic term, that holds a wormhole throat open violates the same energy condition as the negative-mass matter evolved here, though with different field equations (Sec.~\ref{sec:model}). Bondi's self-accelerating pair is itself a standing objection to all such matter, since a pair that accelerates without an external force looks like a perpetual-motion machine~\cite{hoffmann1966,martins1980}. That objection has been examined for point particles and prescribed metrics. The existing nonlinear evolutions of matter that violates energy conditions concern ghost-scalar wormholes, in spherical symmetry~\cite{shinkai2002,gonzalez2009} and in $3{+}1$ dimensions, alone~\cite{shirokov2026,shirokov_spin2026} and in binaries~\cite{shirokov_merger2026}; static stars of phantom fields~\cite{dzhunushaliev2014}; and the collapse of an Alcubierre warp bubble in $3{+}1$ dimensions~\cite{clough2024}. Negative-mass bubbles in de Sitter space~\cite{mbarek2014} and negative-mass black holes~\cite{mann1997} have also been studied. Here we test whether the self-acceleration survives full nonlinear gravity, with finite bodies, backreaction and radiation, for a self-gravitating pair of positive and negative active mass with dynamical, constraint-solved matter.

The question also has observational interest. Negative-mass binaries have been proposed as sources of anomalous gravitational-wave signals, such as anti-chirps, dispersal and runaway motion, which detectors could use to exclude them~\cite{trivedi_loeb2026}. Nojiri and Odintsov~\cite{nojiri2026} find that a bound positive--negative system can form around a compact mass in a cosmological fluid with a negative cosmological constant. We study the complementary case of an isolated, asymptotically flat pair.

In this paper we realize Bondi's sign structure with two complex scalar fields that obey the same field equation, one of which enters the Einstein equations with a negative sign (Sec.~\ref{sec:model}). Both fields form solitonic boson stars, bound by their own self-interaction, which we construct as constraint-satisfying initial data (Sec.~\ref{sec:initialdata}). We evolve them with \texttt{GRTeclyn}~\cite{grteclyn}, the GPU port of \texttt{GRChombo}~\cite{grchombo} built on AMReX~\cite{amrex}, in 24 simulations that include same-sign pairs and isolated stars as controls. The simulations cover head-on pairs released from rest in the weak-field regime, evolved to $t=200$ at up to three resolutions and, for the pair and one isolated star, to $t=1000$. Section~\ref{sec:setup} gives the numerical methods, the diagnostics and the five predictions of Bondi's argument that we test; Sec.~\ref{sec:results} gives the results, Sec.~\ref{sec:weyl} the gravitational radiation, Sec.~\ref{sec:caveats} the systematic checks, and Sec.~\ref{sec:discussion} our conclusions.

\section{The model}
\label{sec:model}

The matter must meet three requirements: (i)~positive inertial and passive mass with negative active mass; (ii)~stable, localized stars; and (iii)~a binding force other than gravity for the negative-mass star, whose own gravity pushes it apart. The obvious candidates each fail one of them. The negative-mass Schwarzschild spacetime is a naked singularity whose linear perturbations grow exponentially~\cite{gleiser2006}, and it contains no matter from which to build initial data. A fluid of negative energy density has no equilibrium of its own: Bonnor's static negative-mass spheres~\cite{bonnor1989} require an inverted interior structure that no non-gravitational force supports. A ghost field (a phantom field in the language of cosmology~\cite{caldwell2002}) carries the sign of its kinetic term into its own field equation. This inverts its effective potential, which unbinds the star, and it makes the inertial and passive masses negative together with the active mass, as for Forward's negative matter~\cite{forward1990}. Wormholes held open by such fields collapse or explode in evolutions, alone~\cite{shinkai2002,gonzalez2009,shirokov2026} and in binaries~\cite{shirokov_merger2026}, and rotation slows this instability~\cite{shirokov_spin2026}. We instead reverse the overall sign of one field's Lagrangian, Eq.~\eqref{eq:action} below, which leaves its field equation unchanged and reverses only its stress-energy tensor.

The stars are solitons: matter bound by its own self-interaction, with gravity a small correction rather than the binding force. We use two independent complex scalar fields, $\Phi_+$ and $\Phi_-$, on one spacetime, with the same sextic potential as in studies of Q-balls~\cite{friedberg1976,coleman1985} and solitonic boson stars~\cite{friedberg1987,evstafyeva2026},
\begin{equation}
V(|\Phi|^2)=\tfrac12 m^2|\Phi|^2-\tfrac14\lambda|\Phi|^4+\tfrac16\mu|\Phi|^6 ,
\label{eq:potential}
\end{equation}
where $m$ is the scalar mass and $\lambda,\mu>0$ are the quartic and sextic couplings: the mass term makes the field oscillate, the attractive quartic term binds it into a star, and the repulsive sextic term prevents the star's collapse. Because the two fields share one potential, their matter physics is identical, and the sign introduced next is the only difference between them. We use units with $G=c=1$ and scalar mass $m=1$.

The model is defined by the action
\begin{equation}
S=\int d^4x\,\sqrt{-g}\left[\frac{{}^{(4)}\!R}{16\pi}+\mathcal{L}[\Phi_+]-\mathcal{L}[\Phi_-]\right],
\label{eq:action}
\end{equation}
where $g$ is the determinant of the spacetime metric, ${}^{(4)}\!R$ its Ricci scalar, and $\mathcal{L}[\Phi]=-\tfrac12\nabla_\mu\Phi^*\nabla^\mu\Phi-V(|\Phi|^2)$. Varying Eq.~\eqref{eq:action} with respect to each field gives the same Klein--Gordon equation for both,
\begin{equation}
\Box\,\Phi_\pm=2\,\frac{\partial V}{\partial|\Phi|^2}\,\Phi_\pm ,
\label{eq:kg}
\end{equation}
where $\Box=g^{\mu\nu}\nabla_\mu\nabla_\nu$: the overall sign of a field's Lagrangian cancels from its own Euler--Lagrange equation, and the factor 2 comes from the $\tfrac12$ in the kinetic term. Varying with respect to the metric gives
\begin{equation}
G_{\mu\nu}=8\pi\left(T_{\mu\nu}[\Phi_+]-T_{\mu\nu}[\Phi_-]\right),
\label{eq:einstein}
\end{equation}
where $T_{\mu\nu}[\Phi]$ is the standard stress-energy tensor of a complex scalar field, and the sign of the Lagrangian carries over to the source.

Equations~\eqref{eq:kg} and~\eqref{eq:einstein} are consistent with each other and with the Bianchi identity $\nabla^\mu G_{\mu\nu}=0$. Each field's stress-energy tensor is conserved on solutions of its own field equation~\eqref{eq:kg}, so the signed sum on the right of Eq.~\eqref{eq:einstein} is conserved as well. The negative source $-T_{\mu\nu}[\Phi_-]$ violates the null energy condition (NEC), $T_{\mu\nu}k^\mu k^\nu\ge0$ for every null vector $k^\mu$, wherever $\Phi_-\neq0$. A field that gravitates with negative energy admits vacuum decay through graviton exchange~\cite{carroll2003,cline2004}, so we make no claim about the model's quantum consistency and use it as a classical probe of how gravity would act on such matter, as in phantom cosmology~\cite{caldwell2002}. The sign enters the evolution equations, which we write in the conformal and covariant Z4 (CCZ4) formulation (Sec.~\ref{sec:setup}), only through algebraic source terms, so their principal part is that of CCZ4 with ordinary scalar matter.

The field $\Phi_-$ therefore moves like ordinary matter, with positive inertial and passive mass, but sources gravity with a negative sign, so its active mass is negative, and so is the Arnowitt--Deser--Misner (ADM) mass of an isolated $\Phi_-$ star. We call the stars of $\Phi_+$ and $\Phi_-$ positive-mass and negative-mass stars; a mixed pair has one star of each field, a same-sign pair two stars of one field. We use the $3{+}1$ split with lapse $\alpha$, shift $\beta^i$ and spatial metric $\gamma_{ij}$, and $n^\mu$ the unit normal to the spatial slices. The Einstein equations are then sourced by the energy density $\rho=n^\mu n^\nu T_{\mu\nu}$, the momentum density $S_i=-\gamma_i{}^\mu n^\nu T_{\mu\nu}$ and the stress $S_{ij}=\gamma_i{}^\mu\gamma_j{}^\nu T_{\mu\nu}$, the projections of the total stress-energy tensor measured by observers moving along $n^\mu$. Each field contributes with its sign $\epsilon_\pm=\pm1$,
\begin{equation}
\rho=\sum_{s=\pm}\epsilon_s\,\rho[\Phi_s],\qquad
S_i=\sum_{s=\pm}\epsilon_s\,S_i[\Phi_s],
\label{eq:adm_sources}
\end{equation}
and likewise for $S_{ij}$, where each bracket is the standard expression for one complex scalar field.

The action~\eqref{eq:action} has no term that couples $\Phi_+$ to $\Phi_-$, so stars of different fields interact only through the metric: in the mixed pair, gravity is the only interaction by construction. Two stars of the same field also attract each other directly through the field's self-interaction where they overlap. This short-range attraction does not depend on the sign of the mass; Sec.~\ref{sec:nulls} measures it, and the mixed pair cannot have it.

\section{Initial data}
\label{sec:initialdata}

\subsection{Isolated stars}

A star of this model is a solution in which the field oscillates with a single frequency, $\Phi=\varphi_0(r)e^{-i\omega t}$; the frequency $\omega$ labels the family of solutions, with $\omega=m$ the free-field limit. The ratio $\lambda^2/\mu$ sets the lowest frequency a star can have, $\omega_{\min}=\sqrt{1-3\lambda^2/16\mu}$~\cite{coleman1985}. All our stars use $\lambda^2/\mu=4.8$ (Table~\ref{tab:params}), so $\omega_{\min}=0.316$.

A star built in flat space violates the constraints once its own gravity is included, so we solve for each star together with its gravity, as for boson stars~\cite{kaup1968,ruffini1969,liebling2023} with a sextic potential, for both signs of the source. We take a conformally flat spatial metric, $\gamma_{ij}=\psi^4\delta_{ij}$, where $\delta_{ij}$ is the flat metric and $\psi$ the conformal factor, and maximal slicing, $K=0$, where $K$ is the trace of the extrinsic curvature. Equations~\eqref{eq:kg} and~\eqref{eq:einstein} then reduce to ordinary differential equations for $\psi(r)$, $\alpha(r)$ and $\varphi_0(r)$. The negative-mass star solves the same equations with the sign of the matter source reversed in the metric equations only. Its gravity is repulsive: $\psi<1$ and $\alpha>1$ in its core. We solve the equations by shooting at fixed frequency, bisecting in the central amplitude, and iterate until the frequency measured at infinity, $\omega/\alpha_\infty$ with $\alpha_\infty$ the asymptotic lapse, matches the requested value to ${\sim}10^{-5}$. Outside $r\approx5.2$ the tabulated solution joins an analytic exterior, the isotropic Schwarzschild metric of the star's ADM mass, with the lapse matched in value and slope at the junction. The field's tail beyond the junction, which carries $1.9\%$ of the ADM mass, is restored by the constraint solve of Sec.~\ref{sec:construction}. The join must be smooth because the lapse does not enter the constraints at $t=0$: a step in it passes every initial-data check, yet the first evolution step turns it into a momentum-constraint violation that grows with resolution.

The reference pair consists of a positive-mass star at $\omega=0.750$ and a negative-mass star at $\omega=0.7603$ (Table~\ref{tab:params}), whose ADM masses $M_\pm$ satisfy $|M_-|=M_+$ to $0.4\%$ (at equal $\omega$ the negative-mass star is heavier in magnitude); we call such a pair an equal-mass pair. Equal ADM masses are the relevant match because, in Bondi's point-mass limit, each body acts on the other through its exterior field, which the ADM mass sets. We write $\bar M=\tfrac12(M_++|M_-|)$ for the mean mass of a pair. The positive-mass star has a root-mean-square (rms) radius $r_{\rm rms}=4.34$, the rms distance from its barycentre weighted by the field's weight $w_s$ of Eq.~\eqref{eq:bary}. Its compactness, $M_+/r_{\rm rms}\sim3\times10^{-3}$, is far from that of horizon formation. Evolved in isolation to $t=200$ (the simulations P and M of Table~\ref{tab:matrix}), each star of the reference pair ends with its peak field amplitude within $0.15\%$ and its central lapse within $7\times10^{-4}$ of the initial values. Each star's conjugate momentum, $\Pi\equiv n^\mu\partial_\mu\Phi=\alpha^{-1}(\partial_t-\beta^i\partial_i)\Phi$, is set from the lapse of its own solution: at $t=0$ the field is real, $\Phi=\varphi_0(r)$, and the momentum is purely imaginary, $\Pi=-(i\omega/\alpha(r))\,\varphi_0(r)$, so that the star starts in its stationary state.

\subsection{The pair}
\label{sec:construction}

We place the two stars at $x=\pm d_0/2$ about the centre of the domain, where $d_0$ is the initial separation, and superpose their fields. \texttt{GRTresna}~\cite{grtresna} then solves the Hamiltonian and momentum constraints for the pair, with each field's sources entering with its sign. Two choices in this step matter for the measurement.

\emph{Both stars on the same slicing.} \texttt{GRTresna} implements the CTTK method~\cite{cttk}, a variant of the conformal transverse-traceless (CTT) decomposition that can solve the Hamiltonian constraint algebraically for the trace $K$, with $K\propto\sqrt{\rho}$. This is imaginary where $\rho<0$, so the negative-mass star requires the maximal-slicing option, $K=0$, which solves for $\psi$ instead. We use $K=0$ for both stars, so that the two constructions differ only in the sign of $\rho$ and the evolution starts without an initial transient in $K$ ($\max|K|\approx10^{-5}$). Both fields are real and both momenta purely imaginary at $t=0$, so the momentum density $S_i$ vanishes identically on the initial slice and the momentum constraint holds from the start.

\emph{The same grid spacing for the solve and the evolution.} The elliptic solve runs on a domain twice the size of the evolution domain. If its grid spacing differed from the evolution's grid spacing $\Delta x$, the metric interpolated onto the evolution grid would be displaced from the scalar field by a fraction of a cell, so each star would start away from the centre of its own gravitational potential and drift, at every resolution, by as much as the effect we measure. We therefore solve at the evolution's $\Delta x$, so that the cell centres coincide and the metric is copied cell by cell onto the evolution grid, out to the outermost layer of cells on every face. The scalar field is set on that grid from each star's radial profile, with its conjugate momentum evaluated in the superposed lapse $\alpha=1+\sum_{s=\pm}(\alpha_s-1)$. The offset between the centroids of the metric perturbation and of the matter is then below $10^{-4}$, the precision of that measurement, in every simulation. The elliptic solver and this step evaluate the matter identically, down to the interpolant of the field and the lapse that enters $\Pi$, since any mismatch would appear as a constraint violation at $t=0$ that no resolution removes. The initial data then enter the evolution at the elliptic solve's own order: the Hamiltonian norm within $r<24$ of the centre of the domain falls at $t=0$ as $\Delta x^{2.0}$ (Fig.~\ref{fig:constraints}a).

Both stars are given the same internal phase. For the mixed pair the relative phase has no effect, because the two fields do not couple. For two stars of the same field it matters: two Q-balls in phase attract each other through the shared field, and two in antiphase repel~\cite{battye_sutcliffe2000,palenzuela2008}. Our same-sign pairs therefore start in the attracting configuration, consistent with the mergers of Sec.~\ref{sec:nulls}. Superposed data for two stars of one field also carry phase-dependent constraint errors, which the constraint solve reduces but does not remove~\cite{helfer2022}; we use these pairs only as controls for the centroid motion.

\section{Numerical methods, diagnostics and tests}
\label{sec:setup}

\subsection{Evolution}

\begin{table}[t]
\caption{Model and numerical parameters, in units $G=c=m=1$.}
\label{tab:params}
\begin{ruledtabular}
\begin{tabular}{lp{0.66\columnwidth}}
Couplings & $\lambda=10240$, $\mu=2.18\times10^{7}$ ($\lambda^2/\mu=4.8$) \\
Reference stars & $\omega=0.750$, $M_+=+0.014350$; $\omega=0.7603$, $M_-=-0.014293$; $\bar M=0.014322$; $r_{\rm rms}=4.34$; central field amplitude $0.019$ \\
Formulation & CCZ4~\cite{alic2012}, damping parameters $(\kappa_1,\kappa_2,\kappa_3)=(3,0,1)$ \\
Gauge & $1{+}\log$ slicing~\cite{bona1995}; Gamma-driver shift~\cite{campanelli2006,baker2006} with damping $\eta=1$ \\
Discretization & fourth-order finite differences; fourth-order Runge--Kutta, $\Delta t=0.02\,\Delta x$ \\
Dissipation & Kreiss--Oliger, coefficient $\sigma=2$ \\
Grid & uniform cube of side $L=64$; $N=128$, 192, 256 cells per side ($\Delta x=0.50$, 0.33, 0.25) \\
Boundaries & Sommerfeld; absorbing layer at $24\le r\le32$ \\
Initial data & \texttt{GRTresna}~\cite{grtresna}, CTTK~\cite{cttk} with $K=0$, on a domain of side $2L$ with the same $\Delta x$ \\
\end{tabular}
\end{ruledtabular}
\end{table}

We evolve the Einstein equations in the CCZ4 formulation~\cite{alic2012} with the gauge, discretization and dissipation of Table~\ref{tab:params}. The CCZ4 variable $\chi=(\det\gamma_{ij})^{-1/3}$, equal to $\psi^{-4}$ on the conformally flat initial data, is below 1 in a positive-mass star and above 1 in a negative-mass star. The time step of Table~\ref{tab:params}, an order of magnitude below the usual CCZ4 value, is set by the stability of the sextic self-interaction: at $\Delta t=0.2\,\Delta x$ the star disperses. Outgoing waves of a massive field reflect from the Sommerfeld outer boundary conditions, which are designed for massless waves, so an absorbing layer, extra Kreiss--Oliger dissipation that rises with radius, covers $24\le r\le32$. With it, the net scalar-field energy flux into the domain, measured on a sphere near the boundary, stays below $10^{-11}$ for the reference pair and below $6\times10^{-6}$ for the same-sign pair PP10, whose merger sheds field.

One implementation pitfall is worth recording for codes built on AMReX. Ghost cells must be filled once per time step, before the first Runge--Kutta stage. The ghost-cell fill of AMReX, \texttt{FillPatch}, replaces a stage's input with the two stored time levels interpolated linearly to the stage time, so refilling at every stage silently reduces the scheme to first order in $\Delta t$ and slowly amplifies each star's oscillating field. No single run shows the defect; it appears only as constraint norms that converge at first order instead of third to fourth (Appendix~\ref{app:constraints}).

All simulations use a uniform grid without adaptive mesh refinement (AMR), so a change of resolution is a change of the grid spacing alone. The three resolutions of Table~\ref{tab:params} put about 9, 13 and 17 grid points across the rms radius of a star. One simulation doubles the side to $L=128$ at $N=256$ ($\Delta x=0.50$), with the absorbing layer at $48\le r\le60$, to extract gravitational waves at larger radii. The reference initial separation is $d_0=10$, about twice the radius that contains $90\%$ of a star's weight $w_s$; other pairs use $d_0=8$ to $20$. Every pair is released from rest, so any motion is dynamical. Most simulations end at $t=200$; Table~\ref{tab:matrix} lists the exceptions. Each simulation ran on one NVIDIA H100 GPU; a simulation to $t=200$ takes about 1, 4 and 11 hours at $N=128$, $192$ and $256$.

\subsection{Diagnostics}

We track each field in two ways. Its barycentre
\begin{equation}
\bar x^{(s)}=\frac{\int x\,w_s\,d^3x}{\int w_s\,d^3x},\qquad w_s=\left(|\Phi_s|^2+|\Pi_s|^2\right)^{1/2},
\label{eq:bary}
\end{equation}
with $\Pi_s$ the conjugate momentum of $\Phi_s$, is computed over the whole domain. Its core position is the $w_s^4$-weighted centroid of the cells where $w_s$ exceeds a quarter of its maximum, which ignores the low-amplitude field a star sheds. The main diagnostic is the midpoint of the two stars and its displacement,
\begin{equation}
X=\tfrac12\left(\bar x^{(+)}+\bar x^{(-)}\right),\qquad \Delta X(t)=X(t)-X(0),
\label{eq:mid}
\end{equation}
which Bondi's argument predicts to accelerate. For a same-sign pair, the centroid of the pair is the barycentre of its single field, Eq.~\eqref{eq:bary}; by symmetry it stays at rest. In the $L=64$ domain the two ways of tracking agree on $\Delta X$ of the mixed pair to $1\%$ (Sec.~\ref{sec:othernum} discusses the doubled domain). The separation $d$ is the coordinate distance between the two cores; we also measure it as a proper distance along the axis, $\int\sqrt{\gamma_{xx}}\,dx$, and as the distance between the two barycentres, which take in the low-amplitude field a star sheds. The acceleration $a$ is twice the quadratic coefficient of the least-squares fit
\begin{equation}
X(t)=X_0+v_0\,t+\tfrac12\,a\,t^2,\qquad 5\le t\le t_{\rm end},
\label{eq:fit}
\end{equation}
which leaves out the first five time units while the gauge adjusts. For point masses with $|M_-|=M_+$, each star is accelerated by the other's active mass, and Bondi's argument predicts
\begin{equation}
a=\frac{\bar M}{d^2}.
\label{eq:bondilaw}
\end{equation}
The matter momentum of each field and their signed total are
\begin{equation}
P^{(s)}=\int S_x[\Phi_s]\,\sqrt{\gamma}\,d^3x,\qquad P=P^{(+)}-P^{(-)},
\label{eq:momentum}
\end{equation}
with $\gamma=\det\gamma_{ij}$. We also record the $L_2$ norms of the Hamiltonian and momentum constraints, $\|\mathcal{H}\|_2$ and $\|\mathcal{M}\|_2$, over the domain, the multipoles (of degree $\ell$) of the Newman--Penrose scalar $\Psi_4$, which carries outgoing gravitational waves, on spheres of radius $R$, and the energy conditions. The isolated positive-mass star satisfies the NEC everywhere.

\begin{table*}[t]
\caption{The 24 simulations. Labels name the stars, P for a positive-mass and M for a negative-mass star, followed by the initial separation $d_0$; E labels equal-mass pairs by $\bar M$ relative to PM20 and R unequal-mass pairs by $|M_-|/M_+$. Unless stated, a simulation uses the stars of the reference pair and the domain $L=64$.}
\label{tab:matrix}
\begin{ruledtabular}
\begin{tabular}{lp{0.29\textwidth}lll}
Label & Configuration & $N$ & $t_{\rm end}$ & Purpose \\
\hline
PM10 & mixed pair, $d_0=10$ & 128, 192, 256 & 200 & reference; resolution study \\
PM8, PM12, PM16, PM20 & mixed pair, $d_0=8$, 12, 16, 20 & 128 & 200 & dependence on separation \\
E0.64, E1.34, E1.57 & as PM20, both stars constructed anew at $(\omega_+,\omega_-)=(0.850,0.859)$, $(0.700,0.710)$, $(0.675,0.686)$, with $|M_-|=M_+$ (to $0.02\%$) & 128 & 200 & dependence on mass \\
R0.60, R0.80, R1.33 & as PM10, one star constructed anew ($\omega_-=0.880$; $\omega_-=0.804$; $\omega_+=0.810$), so that $|M_-|/M_+=0.60$, 0.80, 1.33 & 128 & 200 & unequal masses \\
MP10 & as PM10, the two stars exchanged & 128 & 200 & reversal \\
PM10-t400 & as PM10 & 128 & 400 & the pair beyond $t=200$ \\
PM10-rec & as PM10, domain recentred on the pair during the run & 192 & 1000 & the pair to $t=1000$ \\
PM10-L128 & as PM10, $L=128$ ($\Delta x=0.50$) & 256 & 200 & domain size; wave extraction \\
\hline
PP10 & two positive-mass stars, $d_0=10$ & 128 & 200 & same-sign control \\
MM10 & two negative-mass stars, $d_0=10$ & 128 & 200 & same-sign control \\
P & isolated positive-mass star at the centre of the domain & 128 & 200 & isolated control \\
M & isolated negative-mass star, 5 units off centre & 128, 192, 256 & 200 & isolated control; resolution study \\
M-t1000 & as M & 192 & 1000 & long-term stability \\
\end{tabular}
\end{ruledtabular}
\end{table*}

\subsection{Tests}
\label{sec:tests}

Bondi's argument makes five predictions for a mixed pair with $|M_-|=M_+$, and each has a control:
(i)~the midpoint $X$ accelerates at a constant rate while the separation $d$ stays constant;
(ii)~the signed matter momentum $P$ stays zero while the momentum of each field grows;
(iii)~the acceleration follows Eq.~\eqref{eq:bondilaw}, $a\propto d^{-2}$ at fixed mass and $a\propto\bar M$ at fixed separation, and each star responds to its partner's mass;
(iv)~exchanging the two stars reverses the acceleration;
(v)~configurations without a mass dipole, the same-sign pairs and the isolated stars, stay at rest: neither the centroid of a same-sign pair nor the core of an isolated star moves.
A numerical artefact would fail at least one of these tests: it need not reverse when the stars are exchanged, need not follow $d^{-2}$ or the partner's mass, and would also appear in the controls.

\begin{figure*}[t]
\centering
\includegraphics{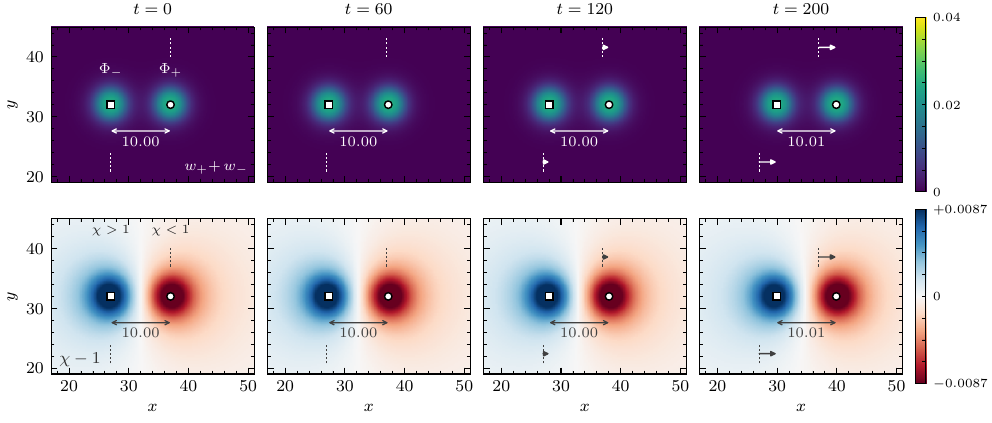}
\caption{The reference simulation (PM10, $N=256$) in the plane $z=0$ at $t=0$, 60, 120 and 200. Top: the sum of the weights $w_s$ of Eq.~\eqref{eq:bary}. Bottom: $\chi-1$, positive for the negative-mass star (left, blue) and negative for the positive-mass star (right, red), on colour scales fixed across snapshots. Circles and squares mark the cores of $\Phi_+$ and $\Phi_-$, dashed ticks the initial positions, arrows the displacement, and the number between the stars their core distance (movie: Video~1, \href{https://youtu.be/zz8_5i7A6zo}{\nolinkurl{youtu.be/zz8_5i7A6zo}}). Both stars move toward $+x$ at constant acceleration while that distance stays constant to $0.05\%$.}
\label{fig:frames}
\end{figure*}

\begin{figure*}[t]
\centering
\includegraphics{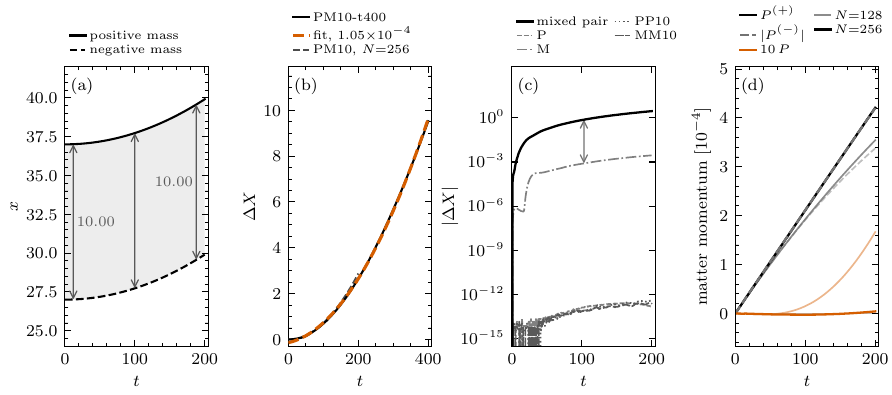}
\caption{(a)~Core positions of the positive-mass (solid) and negative-mass (dashed) star in the reference simulation, with the separation at three times. (b)~$\Delta X$ of PM10-t400 (solid; movie: Video~2, \href{https://youtu.be/n5y8fBBCDUs}{\nolinkurl{youtu.be/n5y8fBBCDUs}}), a constant-acceleration fit over $5\le t\le400$ (orange dashed), and PM10 at $N=256$ (grey). (c)~$|\Delta X|$ of the mixed pair (black) against the core displacement of the isolated stars and the centroid displacement of the same-sign pairs (grey); P, PP10 and MM10 lie in the round-off band, and M, off centre, lies above it. (d)~Matter momentum of each field, $P^{(+)}$ and $P^{(-)}$, and ten times their signed total $P$ (orange), at $N=128$ (light) and $256$ (dark). The pair moves $2.9$ units while, at $N=256$ and $t\ge5$, $P$ stays below $0.5\%$ of either field's momentum.}
\label{fig:traj}
\end{figure*}

\section{Results}
\label{sec:results}

\begin{table*}[t]
\caption{Measured results against Bondi's predictions, by simulation (Table~\ref{tab:matrix}). The $C$ ratio compares a star's fitted acceleration constant $C$, in $\ddot x=\pm C/d^2$, with its value in PM10 (Sec.~\ref{sec:scaling}).}
\label{tab:results}
\begin{ruledtabular}
\begin{tabular}{lp{0.28\textwidth}p{0.22\textwidth}l}
Simulation & Quantity & Measured & Expected \\
\hline
PM10, $N=256$ & $\Delta X(200)$ & $2.87\pm0.03$ & $2.86$ for $a=\bar M/d_0^2$ \\
PM10, $N=256$ / 192 / 128 & $a$ $[10^{-4}]$ & $1.45$ / $1.43$ / $1.27$ & $1.43$ \\
PM10, three resolutions & order of convergence of $\Delta X(200)$; of $a$ & $4.4$; $4.2$ & 4 \\
PM10, $N=256$ & $|P|/P^{(+)}$ at $t=200$ & $0.001$ & 0 \\
PM10, $N=256$ & change of $d$ over $t\le200$: cores; proper & $+0.05\%$; $+0.15\%$ & 0 \\
PM10-t400 & $a$ over $133\le t\le266$; $300\le t\le400$ $[10^{-4}]$ & $1.06$; $0.82$ & constant \\
PM10-rec & speed of the leading; trailing star at $t=1000$ & $0.115$; $0.108$ & equal \\
MP10 & $\Delta X(200)$; $a$, ratio to PM10 at $N=128$ & $-1.0000$; $-1.0001$ & $-1$ \\
PM8--PM20 & exponent of $a(d_0)$ & $-1.95$ & $-2$ \\
PM8, PM12, PM16, PM20 & $a\,d_0^2/\bar M$, ratio to PM10 & 0.994, 1.009, 1.025, 1.036 & 1 \\
E0.64, PM20, E1.34, E1.57 & exponent of $a(\bar M)$ & $0.97\pm0.16$ & 1 \\
E0.64, E1.34, E1.57 & $a\,d_0^2/\bar M$, ratio to PM20 & 1.030, 0.999, 1.001 & 1 \\
R0.60, R0.80, R1.33 & $C$ ratio, partner changed & 0.610, 0.806, 0.751 & 0.600, 0.803, 0.747 \\
R0.60, R0.80, R1.33 & $C$ ratio, partner unchanged & 1.034, 1.006, 0.994 & 1 \\
R0.60, R0.80, PM10, R1.33 & $|M_-|/M_+$ at which $d(200)=d_0$ & $0.99\pm0.02$ & 1 \\
PP10; MM10 & largest centroid displacement & $4\times10^{-13}$; $3\times10^{-13}$ & 0 \\
P; M at $N=256$ & largest core displacement over $t\le200$ & $3\times10^{-13}$; $2.8\times10^{-3}$ & 0 \\
M-t1000 & change of the peak field over $t\le1000$ & $+0.11\%$ & 0 \\
PM10-L128 & exponent of $|\Psi_4|_{\ell=2}(R)$ & $-4.7$ & $-1$ if radiative \\
PM10-L128 & change of $\Delta X(200)$ from PM10 at $N=128$: cores; barycentres & $+0.3\%$; $-4.8\%$ & 0 \\
\end{tabular}
\end{ruledtabular}
\end{table*}

\subsection{The mixed pair accelerates at constant separation}
\label{sec:runaway}

Does the mixed pair accelerate as a single object (test~i)? In the reference simulation, PM10 at $N=256$, the positive-mass star starts at $x=37$ and the negative-mass star at $x=27$, and both accelerate toward $+x$ (Figs.~\ref{fig:frames} and~\ref{fig:traj}a). The midpoint moves $\Delta X=2.87\pm0.03$ by $t=200$, and a single parabola, Eq.~\eqref{eq:fit}, fits the whole trajectory with $a=(1.45\pm0.02)\times10^{-4}$, with errors from the difference between $N=192$ and $256$ (Sec.~\ref{sec:convergence}). The pair keeps accelerating in the longer run PM10-t400, at $N=128$, and moves at $0.042c$ at $t=400$ (Fig.~\ref{fig:traj}b; Sec.~\ref{sec:latetime}). Section~\ref{sec:t1000} follows the pair to $t=1000$. Over $t\le200$ the separation between the cores stays constant to $0.05\%$ in coordinate distance and to $0.15\%$ in proper distance (Table~\ref{tab:results}). The distance between the barycentres dips by $2.5\%$ near $t\approx50$ and ends $0.9\%$ below $d_0$, because each star sheds low-amplitude field, which the whole-domain barycentres take in; the rms radius of each field grows by $16$ to $22\%$ by $t=200$.

The stars keep their structure while they accelerate. Their peak field amplitude changes by less than $0.3\%$, the minimum lapse stays above $0.99$, and the minimum of $\chi$ stays above $0.98$, far from collapse, which would drive $\chi$ toward zero.

\subsection{The signed momentum stays zero}
\label{sec:momentum}

Does the total momentum stay zero while the stars move (test~ii)? The matter momentum of each field grows steadily, while the signed total $P$ of Eq.~\eqref{eq:momentum} stays below $0.5\%$ of either term at $N=256$ once the gauge has adjusted, $t\ge5$ (Fig.~\ref{fig:traj}d and Table~\ref{tab:results}). At $t=200$ this residual falls by a factor of $49$ from $N=128$ to $256$, consistent with a discretization error. $P$ is the matter momentum rather than the ADM momentum; we expect the difference, the momentum carried by the gravitational field, to be small at compactness $3\times10^{-3}$.

\subsection{The acceleration follows $a=\bar M/d^2$}
\label{sec:forcelaw}

\begin{figure}[t]
\centering
\includegraphics{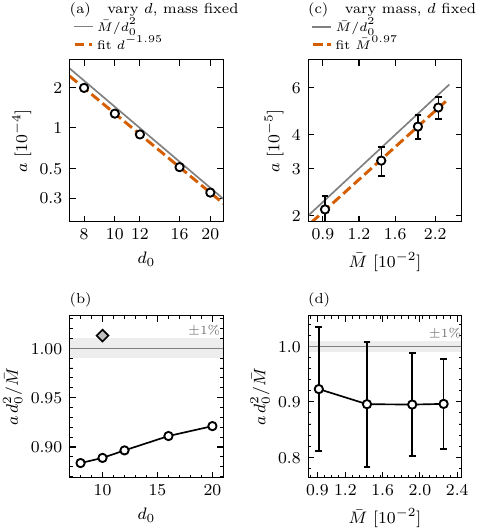}
\caption{(a)~Acceleration $a$ against initial separation $d_0$ at $N=128$ (circles), with a power-law fit (orange dashed) and $\bar M/d_0^2$ (grey). (b)~$a\,d_0^2/\bar M$ against $d_0$; the diamond is the $N=256$ value at $d_0=10$ and the band is $\pm1\%$. (c)~$a$ against $\bar M$ for the four equal-mass pairs at $d_0=20$, with the same fit and reference curves; (d)~their $a\,d_0^2/\bar M$. Error bars are half the spread of fits over four disjoint time windows. The acceleration scales as $d^{-1.95}$ and as $\bar M^{0.97}$; the $N=128$ values sit $8$ to $12\%$ below $\bar M/d^2$, the truncation error of Sec.~\ref{sec:convergence}, which cancels from exponents and ratios.}
\label{fig:forcelaw}
\end{figure}

Does the acceleration follow Eq.~\eqref{eq:bondilaw} (test~iii)? Five pairs with $d_0=8$ to $20$ test the dependence on separation (Fig.~\ref{fig:forcelaw}a,b). A power law through their accelerations gives $a\propto d^{-1.95}$. The ratio $a\,d_0^2/\bar M$ rises by $4\%$ from $d_0=8$ to $20$ (Fig.~\ref{fig:forcelaw}b), and this trend accounts for the deviation of the exponent from $-2$. The signed momentum $P$ at $t=200$ falls with separation as the acceleration does.

These five simulations ran at $N=128$, whose acceleration at $d_0=10$ lies $12\%$ below the $N=256$ value. A truncation error common to the five pairs cancels from the exponent and from every ratio in Sec.~\ref{sec:scaling}; the $4\%$ trend of $a\,d_0^2/\bar M$ bounds what does not cancel, together with the finite-size corrections. The absolute scale is set at $N=256$: the reference pair gives $a\,d_0^2/\bar M=1.01\pm0.01$, so at a separation of two stellar diameters Eq.~\eqref{eq:bondilaw} holds to about $1\%$.

Four equal-mass pairs at $d_0=20$ test the dependence on mass (Fig.~\ref{fig:forcelaw}c,d). In each pair both stars were constructed at new frequencies so that $|M_-|=M_+$, since an unequal pair changes its separation (Sec.~\ref{sec:scaling}) and Eq.~\eqref{eq:fit} then no longer applies. Over a factor $2.46$ in $\bar M$ the accelerations give $a\propto\bar M^{0.97\pm0.16}$, and each pair's $a\,d_0^2/\bar M$ lies within $3\%$ of the PM20 value. The error bars are the spread over fit windows (Sec.~\ref{sec:othernum}). These values use the cores; the barycentres agree except for the lightest pair, which is the most diffuse and sheds the most field.

\subsection{Exchanging the stars reverses the acceleration}
\label{sec:mirror}

Does the motion reverse when the two stars are exchanged (test~iv)? MP10 is PM10 at $N=128$ with the two stars swapped in place. Its $\Delta X(200)$ and its acceleration are those of PM10 with the opposite sign, to three parts in $10^5$ and one part in $10^4$ (Table~\ref{tab:results}). This excludes any asymmetry of the numerical method that does not depend on which star carries the negative mass, such as a biased stencil, an asymmetric domain or a direction-dependent diagnostic. It does not exclude an artefact sourced by the matter, which would reverse as well, but tests~(iii) and~(v) exclude it.

\subsection{Pairs without a mass dipole stay at rest}
\label{sec:nulls}

\begin{figure}[t]
\centering
\includegraphics{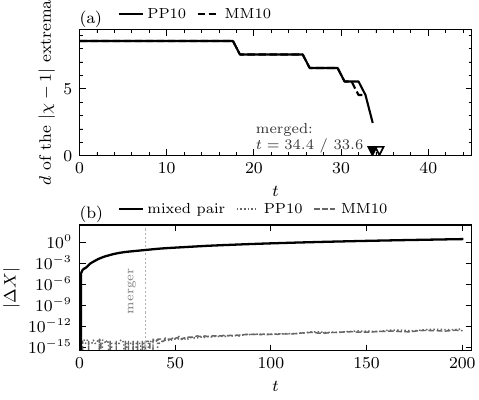}
\caption{The same-sign controls, both at $N=128$ (movies: Videos~3, \href{https://youtu.be/ANS986k4RsA}{\nolinkurl{youtu.be/ANS986k4RsA}}, and~4, \href{https://youtu.be/YKgzOYMcjk4}{\nolinkurl{youtu.be/YKgzOYMcjk4}}). (a)~Distance between the two extrema of $|\chi-1|$ against time for PP10 (solid) and MM10 (dashed); the triangles mark the merger times. (b)~$|\Delta X|$ of the mixed pair and the centroid displacement of PP10 and MM10. Although gravity attracts the stars of PP10 and repels those of MM10, both pairs merge at nearly the same time, through the attraction of their shared field, and their centroids stay at round-off.}
\label{fig:samesign}
\end{figure}

Do pairs without a mass dipole stay at rest (test~v)? The isolated positive-mass star P sits at the centre of the domain, where the configuration is symmetric under reflection, and its core stays at its initial position to round-off, below $10^{-12}$, over $t\le200$ (Fig.~\ref{fig:traj}c). The isolated negative-mass star M starts five units off centre, where no symmetry holds it in place; at $N=256$ its core moves by at most $2.8\times10^{-3}$ on any axis, a thousand times less than $\Delta X$. The barycentre of M, the quantity that corresponds to the barycentres used for $\Delta X$, moves $1.1\times10^{-2}$, still about $260$ times less; it takes in the low-amplitude field the star sheds, and its motion decreases with resolution (Fig.~\ref{fig:convergence_lone}).

The same-sign pairs merge, PP10 at $t=34.4$ and MM10 at $t=33.6$ (Fig.~\ref{fig:samesign}a). Gravity attracts the two positive masses and repels the two negative ones (Fig.~\ref{fig:schematic}), so if gravity set the timescale, PP10 would merge while MM10 flew apart. The stars are instead pulled together by their shared field (Sec.~\ref{sec:model}). The required strength is consistent with this: Newtonian free fall from $d=10$ at these masses takes $t\approx207$, so merging at $t\approx34$ requires a force about $36$ times that of gravity. Each same-sign pair is symmetric under reflection through its midpoint, and the evolution preserves that symmetry: through infall, merger and the oscillations that follow, the centroids of both pairs stay at their initial positions to round-off, below $5\times10^{-13}$, twelve orders of magnitude below $\Delta X$ (Fig.~\ref{fig:samesign}b). The merged objects keep the sign of $\chi-1$.

The mixed pair shows no such attraction. At $d_0=10$, where same-field pairs merge within $t\approx34$, the separation of the mixed pair stays constant to $0.05\%$ for $t\le200$ (Table~\ref{tab:results}); even a few percent of a force $36$ times gravity would have changed $d$ by far more. The mixed pair's acceleration also falls as a power of the separation, whereas an overlap force falls exponentially, and it follows the partner's ADM mass (Sec.~\ref{sec:scaling}), on which a force through the field does not depend.

\subsection{Each star follows its partner's mass}
\label{sec:scaling}

\begin{figure}[t]
\centering
\includegraphics{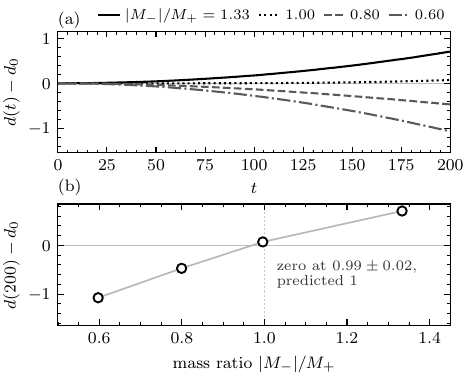}
\caption{(a)~Change of the separation, $d(t)-d_0$, for R0.60, R0.80, PM10 and R1.33 ($d_0=10$). (b)~$d(200)-d_0$ against $|M_-|/M_+$. The separation grows when the negative-mass star is the heavier and shrinks when it is the lighter; the change crosses zero at $|M_-|/M_+=0.99\pm0.02$, as predicted.}
\label{fig:massscale}
\end{figure}

Does each star respond to its partner's mass (test~iii)? In Bondi's argument, changing one star's mass changes its partner's acceleration and leaves its own unchanged. Three unequal-mass pairs at $d_0=10$ test this (Fig.~\ref{fig:massscale}a). We constructed one star of each pair anew, at the frequencies of Table~\ref{tab:matrix}, which gives $|M_-|/M_+=0.60$, $0.80$ and $1.33$. Because the separation now changes, we fit each star's acceleration constant $C$ in $\ddot x=\pm C/d(t)^2$ with the measured $d(t)$, by regressing the core position $x(t)$ on $\{1,t,I_2(t)\}$ for $t\ge5$, where $I_2$ is the double time integral of $d^{-2}$. When a star's partner is changed, its acceleration scales with the partner's mass to within $0.4$ to $1.8\%$; when the partner is unchanged, the acceleration stays within $3.4\%$ of its PM10 value (Table~\ref{tab:results}). No lighter partner exists at these couplings: neither family has a star lighter than about $0.54$ of the reference mass.

Unequal masses also change the separation, in the direction Bondi's argument predicts. When the negative-mass star is the heavier, the positive-mass star in front is pushed away more strongly than the negative-mass star behind it is pulled, and the separation grows; when the negative-mass star is the lighter, the separation shrinks. Across the four pairs, the change of the separation at $t=200$ is monotonic in $|M_-|/M_+$ and crosses zero at $0.99\pm0.02$ (Fig.~\ref{fig:massscale}b), as predicted. The error is the spread between ways of measuring the separation (between the barycentres, between the cores, as a proper distance, and averaged over $190\le t\le200$).

\subsection{Evolution to $t=1000$}
\label{sec:t1000}

Does the acceleration persist, and does the negative-mass star last? Two simulations follow the system five times longer than the main set. PM10-rec evolves the reference pair at $N=192$ with the domain recentred on the pair in steps of two units, accumulating the tracked positions across the shifts (movie: Video~5, \href{https://youtu.be/VeqhSRbtVas}{\nolinkurl{youtu.be/VeqhSRbtVas}}). Both stars accelerate forward throughout the run: by $t=1000$ the pair has moved 63 units, and the leading and trailing stars move at $0.115c$ and $0.108c$ (Table~\ref{tab:results}). The acceleration is not constant over this span: the midpoint value halves, from $1.42\times10^{-4}$ over $5\le t\le200$ to $0.70\times10^{-4}$ over $800\le t\le1000$. The separation opens to $13.5$ at $t=1000$, because the trailing star accelerates $9\%$ less than the leader. Equation~\eqref{eq:bondilaw}, evaluated with the measured late-window separation and matter, predicts $0.87\times10^{-4}$: the opened separation accounts for three quarters of the fall. The remainder, like the opening itself, is truncation error (Sec.~\ref{sec:latetime}): at this resolution the stars lose $2.7\%$ of their matter to dissipation by $t=1000$, while their peak field amplitudes change by only $0.3\%$. The Hamiltonian constraint norm over $r<24$ grows by a factor of $3.9$ up to $t\approx600$ and is flat afterwards; the momentum norm stays flat.

The isolated negative-mass star M-t1000, evolved at the same resolution to $t=1000$, neither collapses nor disperses: its peak field amplitude changes by $0.11\%$, its matter by $2.3\%$, and its core moves $0.12$ toward the centre of the domain, almost all of it after $t=400$. A star of negative ADM mass thus shows no sign of instability over the evolved time.

\section{Gravitational radiation}
\label{sec:weyl}

Does the accelerating pair radiate? The exact solutions for uniformly accelerated pairs are radiative~\cite{bonnor_swami1964,bicak1983}, and dipole radiation has been proposed for negative-mass binaries~\cite{trivedi_loeb2026}. In our model dipole radiation is absent for the same reason as in ordinary general relativity. The source of gravity is the signed stress-energy tensor, which is conserved (Sec.~\ref{sec:model}), so the time derivative of the signed mass dipole equals the signed momentum $P$, which stays zero (Sec.~\ref{sec:momentum}). Independently, $\Psi_4$ has spin weight $-2$, so its multipoles begin at $\ell=2$. The leading candidate is therefore quadrupole radiation from the accelerating pair. For a pair with zero momentum, the signed mass quadrupole along the axis is $Q\simeq2\bar M d\,X(t)$, so at constant separation and acceleration its second time derivative, $\ddot Q=2\bar M a d$, is constant. Computing $Q$ from the tracked positions of the two stars, we find $\ddot Q=4.15\times10^{-5}$, within $0.2\%$ of the predicted $2\bar M a d$ and constant to better than $1\%$ over $20\le t\le190$. The quadrupole luminosity is proportional to $\dddot Q^{\,2}$ and vanishes for constant $\ddot Q$. From the measured bound $|\dddot Q|\le2\times10^{-9}$, the energy radiated over the run is at most of order $10^{-17}$, thirteen orders of magnitude below the pair's total mass, $M_++M_-=5.7\times10^{-5}$.

\begin{figure}[t]
\centering
\includegraphics{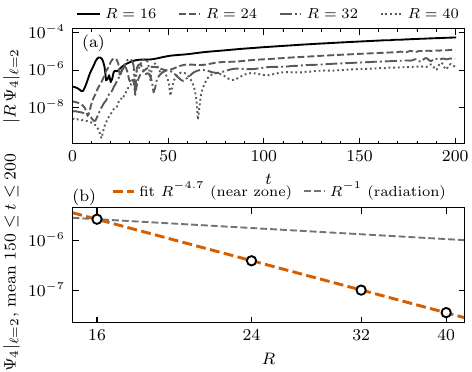}
\caption{(a)~$\ell=2$ amplitude of $R\,\Psi_4$ on spheres of radius $R=16$, 24, 32 and 40 in PM10-L128. (b)~The mean of $|\Psi_4|_{\ell=2}$ over $150\le t\le200$ against $R$, with the fit $\propto R^{-4.7}$ (orange dashed) and the $R^{-1}$ fall-off of radiation (grey). The amplitude falls far faster than radiation would: all four spheres lie in the near zone, and no flux is detected.}
\label{fig:wavezone}
\end{figure}

We extract $\Psi_4$ on spheres at $R=16$ to $40$ in PM10-L128, whose domain is twice as large (Fig.~\ref{fig:wavezone}a). The $\ell=2$ amplitude falls as $R^{-4.7}$ (Fig.~\ref{fig:wavezone}b), and $R\,|\Psi_4|$ drops by a factor 30 across the four spheres, where an outgoing wave would keep it constant. All four spheres therefore lie in the near zone, where $\Psi_4$ reflects the curvature of the source rather than a radiated flux, and we detect no radiation, in agreement with the estimate above. The radiative regime needs much higher speeds than the $0.03c$ the pair reaches by $t=200$.

\section{Convergence and systematic checks}
\label{sec:caveats}

\subsection{Resolution}
\label{sec:convergence}

\begin{figure*}[p]
\centering
\includegraphics{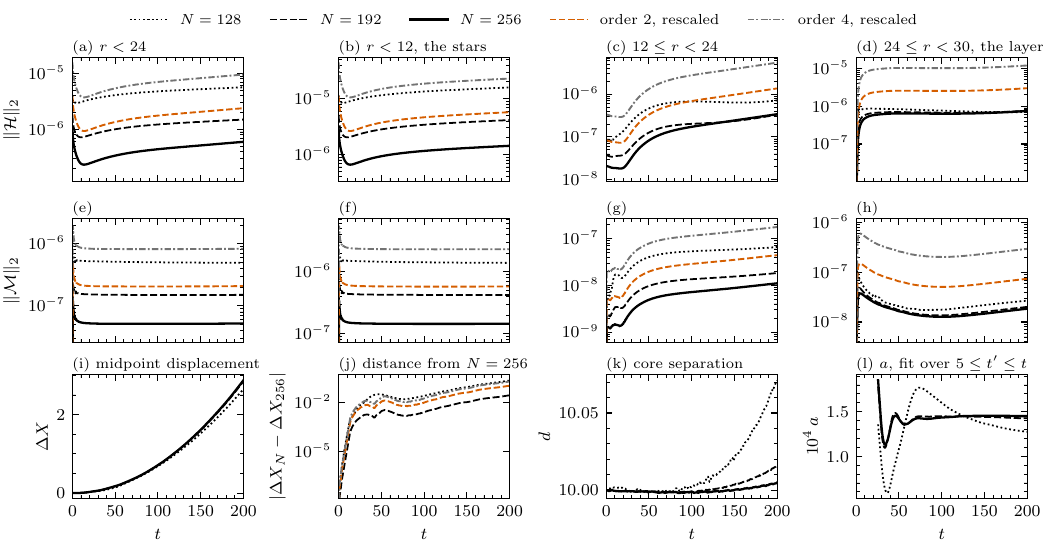}
\caption{Convergence of PM10 at $N=128$ (dotted), 192 (dashed) and 256 (solid). (a)--(d)~$L_2$ norm of the Hamiltonian and (e)--(h)~of the momentum constraint over $r<24$, the stars ($r<12$), $12\le r<24$ and the absorbing layer ($24\le r<30$); orange and grey, the $N=256$ norm multiplied by $2^p$ for $p=2$ and 4, on which the $N=128$ norm lies at order $p$. (i)~$\Delta X$; (j)~$|\Delta X_N-\Delta X_{256}|$, with the $N=192$ curve rescaled so that the $N=128$ curve lies on it at order 2 (orange) or 4 (grey); (k)~the separation $d$ between the cores; (l)~the acceleration $a$ fitted over $5\le t'\le t$. Inside $r<24$ the constraints converge at order 3 to 4 and $\Delta X$ at order 4; in the absorbing layer they do not converge.}\label{fig:constraints}
\vspace{2pt}
\includegraphics{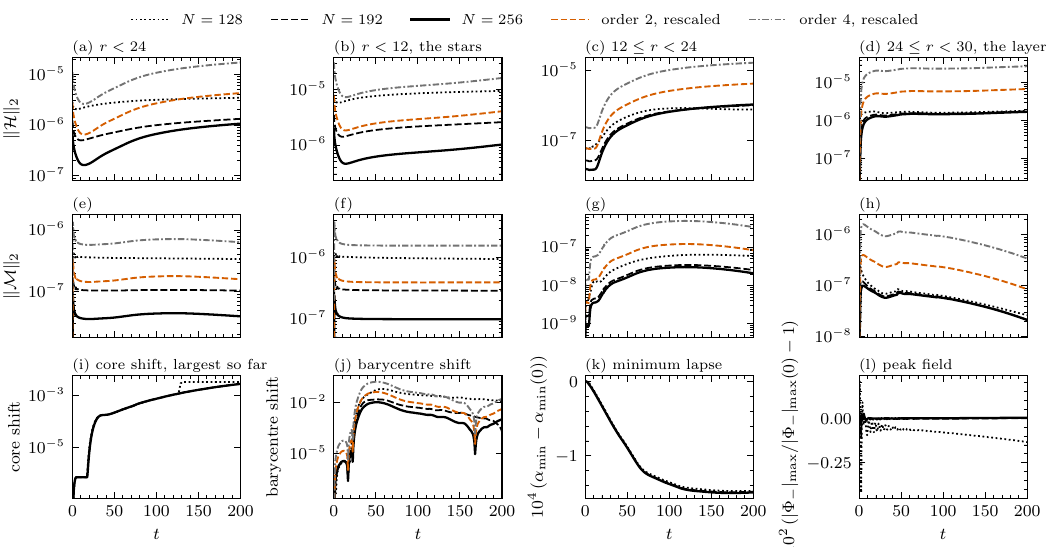}
\caption{Convergence of M, the isolated negative-mass star: (a)--(h)~the constraint norms over the same four regions, with line styles and colours as in Fig.~\ref{fig:constraints}. (i)~Largest displacement of the core on any axis since $t=0$; (j)~displacement of the barycentre, with the $N=256$ curve multiplied by $2^p$ for $p=2$ and 4; (k)~change of the minimum lapse; (l)~relative change of the peak field. Around the star, $r<12$, both constraints converge at order 3 to 4, and the core, the lapse and the peak field agree between $N=192$ and 256.}\label{fig:convergence_lone}
\end{figure*}

We evolved PM10 at $N=128$, 192 and 256 (Fig.~\ref{fig:constraints}i), with grid spacings $h$ in the ratio $6:4:3$, and measured the self-convergence of each headline observable $f$ from its pairwise differences against the value expected at order $n$,
\begin{equation}
Q_{\rm meas}=\frac{f_1-f_2}{f_2-f_3},\qquad
Q_n=\frac{h_1^n-h_2^n}{h_2^n-h_3^n}.
\label{eq:qmeas}
\end{equation}
For $\Delta X(200)=2.648$, $2.846$ and $2.874$ the differences have the same sign and $Q_{\rm meas}=6.9$ lies between $Q_4=5.9$ and $Q_5=8.7$, so $\Delta X$ converges at order $4.4$; the acceleration converges at order $4.2$. We quote the $N=256$ values with the $192$--$256$ difference as the error bar; Richardson extrapolation at the measured order gives $\Delta X(200)=2.89$ and $a=1.46\times10^{-4}$, inside those error bars. The lowest resolution lies $8\%$ below in $\Delta X(200)$. The Hamiltonian constraint inside $r<24$ converges at order $3.3$ late in the run; Appendix~\ref{app:constraints} shows the constraints and the observables at the three resolutions, by region.

\subsection{Domain size and fit window}
\label{sec:othernum}

Doubling the side of the domain to $L=128$ at the grid spacing of $N=128$ changes $\Delta X(200)$ measured from the cores by $+0.3\%$ and leaves the separation unchanged, so the outer boundary does not drive the motion. Measured from the whole-domain barycentres, $\Delta X(200)$ changes by $-4.8\%$: the doubled domain keeps the shed field that the absorbing layer of the $L=64$ domain removes, and the barycentres take it in. Every other quoted $\Delta X$ is measured in the $L=64$ domain. Fitting $a$ over subwindows of a run instead of $5\le t\le t_{\rm end}$ changes it by up to $10\%$ at $N=128$, where the stars slowly lose matter (Sec.~\ref{sec:latetime}); this window spread sets the error bars of Fig.~\ref{fig:forcelaw} and largely cancels from ratios of simulations fitted over the same windows.

\subsection{Gauge}

$\Delta X$ and $d$ are coordinate quantities, so we checked the gauge directly. The shift at the two cores is nearly antisymmetric, so its mean, the only part that could move the pair as a whole, stays below $8\times10^{-6}$, against a midpoint velocity of $3\times10^{-2}$. Integrated over the run, it accounts for $0.011\%$ of $\Delta X$. The proper distance between the cores, independent of the spatial coordinates along the axis, changes by only $0.15\%$ (Table~\ref{tab:results}).

\subsection{The pair beyond $t=200$}
\label{sec:latetime}

At $N=128$ the separation grows after $t\approx200$, to $10.84$ at $t=400$ in PM10-t400, and the fitted acceleration falls by a quarter between the two windows of Table~\ref{tab:results}. We attribute both effects to truncation error; Sec.~\ref{sec:t1000} links the two through Eq.~\eqref{eq:bondilaw}. At this resolution the stars lose $5.9\%$ of their matter to dissipation by $t=400$ ($2.6\%$ by $t=200$, against less than $1\%$ at $N=192$ and 256), and the separation change converges away: the change of the core separation at $t=200$ falls fourteenfold from $N=128$ to 256, for a factor two in grid spacing. With equal masses both stars accelerate equally at every separation, so nothing restores the separation, and any small difference between the two accelerations accumulates. We therefore quote pair quantities for $t\le200$ only.

\section{Conclusions}
\label{sec:discussion}

We evolved Bondi's pair of positive and negative active mass as a full initial-value problem in $3{+}1$ numerical relativity, with two complex scalar fields that differ only in the sign with which they source the Einstein equations. Released from rest, a pair with equal $|M_{\rm ADM}|$ accelerates as a single object, at a constant rate and constant separation over the converged span $t\le200$, while its total signed momentum stays zero. The acceleration follows Bondi's Newtonian law in both separation and mass, each star responds to its partner's mass, and exchanging the two stars reverses the motion. The same-sign pairs and the isolated stars, which have no mass dipole, stay at rest. In the longest run the pair is still accelerating at a tenth of the speed of light.

We detect no gravitational radiation: the signed mass dipole stays constant, the second time derivative of the quadrupole is constant, and every extraction sphere lies in the near zone. The negative-mass star exists as constraint-satisfying, asymptotically flat initial data. It neither collapses nor disperses over 1000 time units, evolved alone or in the accelerating pair, and its peak field amplitude changes by less than $0.3\%$.

These results hold for head-on pairs released from rest, in the weak-field regime (compactness $3\times10^{-3}$), at speeds up to a few percent of the speed of light, and over the evolved times. Two extensions follow. Releases with a transverse offset would test whether the pair is stable against sideways perturbations. Relativistic speeds would test the question that motivates this configuration: the pair's total mass is only $0.4\%$ of either star's, and no positive-mass theorem bounds the energy it could radiate. The speed here is limited by resolution rather than by run length, so that measurement needs finer grids; near $0.3c$ it would provide a concrete source for the signals proposed in Ref.~\cite{trivedi_loeb2026}. Our results do not address whether such matter exists.

\section*{Declarations}
\label{sec:repro}

\textit{Funding.} This work was supported by Gravity Frontiers (\url{https://www.gravityfrontiers.org/en}), which funded the research and the development of the simulation software.

\textit{Competing interests.} The authors have no competing interests to declare.

\textit{Data availability.} The data that support this study are openly available in the repository of Ref.~\cite{repo}, under \texttt{results/bondi-dipole-runaway}, archived on Zenodo (DOI 10.5281/zenodo.23662692). For every simulation they include the time series of every diagnostic and the parameter files, with the script that regenerates every figure and prints every number quoted here. The Supplemental Material contains movies of PM10 (Video~1), PM10-t400 (Video~2), the same-sign controls (Videos~3 and~4) and PM10-rec (Video~5).

\textit{Code availability.} The evolution code (\texttt{GRTeclyn}~\cite{grteclyn,amrex,grchombo}) is openly available in the same repository, and the initial-data code (\texttt{GRTresna}~\cite{grtresna}), with the changes used here, at \url{https://github.com/Nikchik-coder/GRTresna}.

\begin{acknowledgments}
The authors acknowledge the GRTL Collaboration for \texttt{GRChombo}, \texttt{GRTresna},
and \texttt{GRTeclyn}.
\end{acknowledgments}

\appendix

\section{Resolution study}
\label{app:constraints}

Figures~\ref{fig:constraints} and~\ref{fig:convergence_lone} show the resolution study of PM10 and of the isolated star M at $N=128$, 192 and 256. The constraints are $L_2$ norms over regions centred on the domain: the domain is mostly vacuum, so relative norms would be dominated by small denominators. The exact value of each constraint is zero, so multiplying one resolution's norm by $2^p$ and comparing it with another is a valid test of order $p$. Every elliptic solve stops at a relative Hamiltonian residual near $10^{-5}$ at $N=128$; measured on the evolution grid, the $t=0$ norm falls as $\Delta x^{2.0}$ (Sec.~\ref{sec:construction}). During the evolution the constraints converge at order 3 to 4 inside $r<24$ for the pair, and around the star ($r<12$) for M; in the absorbing layer, whose dissipation terms are not part of the field equations, they do not converge, and the layer dominates the whole-domain norm. Panels (i)--(l) of both figures show the observables: $\Delta X$ of the pair converges at order 4, and the core, lapse and peak field of the isolated star agree between $N=192$ and 256. The constraint norms of the same-sign pair PP10 stay bounded through its merger.

\FloatBarrier


\begin{thebibliography}{99}

\bibitem{bondi1957}
H.~Bondi,
\newblock ``Negative mass in general relativity,''
\newblock Rev.\ Mod.\ Phys.\ \textbf{29}, 423 (1957).

\bibitem{bonnor_swami1964}
W.~B.~Bonnor and N.~S.~Swaminarayan,
\newblock ``An exact solution for uniformly accelerated particles in general relativity,''
\newblock Z.\ Phys.\ \textbf{177}, 240 (1964).

\bibitem{bicak1983}
J.~Bi\v{c}\'ak, C.~Hoenselaers, and B.~G.~Schmidt,
\newblock ``The solutions of the Einstein equations for uniformly accelerated particles without nodal singularities. II. Self-accelerating particles,''
\newblock Proc.\ R.\ Soc.\ London A \textbf{390}, 411 (1983); see also \textbf{390}, 397 (1983).

\bibitem{bonnor1989}
W.~B.~Bonnor,
\newblock ``Negative mass in general relativity,''
\newblock Gen.\ Relativ.\ Gravit.\ \textbf{21}, 1143 (1989).

\bibitem{forward1990}
R.~L.~Forward,
\newblock ``Negative matter propulsion,''
\newblock J.\ Propulsion Power \textbf{6}, 28 (1990).

\bibitem{hoffmann1966}
B.~Hoffmann,
\newblock ``Negative mass and the quasars,''
\newblock in \emph{Perspectives in Geometry and Relativity}, edited by B.~Hoffmann (Indiana University Press, Bloomington, 1966), p.~176.

\bibitem{martins1980}
R.~de~A.~Martins,
\newblock ``Causal paradoxes implied by the hypothetical coexistence of positive- and negative-mass matter,''
\newblock Lett.\ Nuovo Cimento \textbf{28}, 265 (1980).

\bibitem{caldwell2002}
R.~R.~Caldwell,
\newblock ``A phantom menace? Cosmological consequences of a dark energy component with super-negative equation of state,''
\newblock Phys.\ Lett.\ B \textbf{545}, 23 (2002).

\bibitem{morris_thorne1988}
M.~S.~Morris and K.~S.~Thorne,
\newblock ``Wormholes in spacetime and their use for interstellar travel: A tool for teaching general relativity,''
\newblock Am.\ J.\ Phys.\ \textbf{56}, 395 (1988).

\bibitem{alcubierre1994}
M.~Alcubierre,
\newblock ``The warp drive: hyper-fast travel within general relativity,''
\newblock Class.\ Quantum Grav.\ \textbf{11}, L73 (1994).

\bibitem{clough2024}
K.~Clough, T.~Dietrich, and S.~Khan,
\newblock ``What no one has seen before: gravitational waveforms from warp drive collapse,''
\newblock Open J.\ Astrophys.\ \textbf{7}, 10.33232/001c.121868 (2024).

\bibitem{mann1997}
R.~B.~Mann,
\newblock ``Black holes of negative mass,''
\newblock Class.\ Quantum Grav.\ \textbf{14}, 2927 (1997).

\bibitem{nojiri2026}
S.~Nojiri and S.~D.~Odintsov,
\newblock ``May negative mass objects exist in the sky?,''
\newblock Phys.\ Dark Univ.\ \textbf{53}, 102364 (2026).

\bibitem{gleiser2006}
R.~J.~Gleiser and G.~Dotti,
\newblock ``Instability of the negative mass Schwarzschild naked singularity,''
\newblock Class.\ Quantum Grav.\ \textbf{23}, 5063 (2006).

\bibitem{shinkai2002}
H.~Shinkai and S.~A.~Hayward,
\newblock ``Fate of the first traversible wormhole: Black-hole collapse or inflationary expansion,''
\newblock Phys.\ Rev.\ D \textbf{66}, 044005 (2002).

\bibitem{friedberg1976}
R.~Friedberg, T.~D.~Lee, and A.~Sirlin,
\newblock ``Class of scalar-field soliton solutions in three space dimensions,''
\newblock Phys.\ Rev.\ D \textbf{13}, 2739 (1976).

\bibitem{coleman1985}
S.~Coleman,
\newblock ``Q-balls,''
\newblock Nucl.\ Phys.\ B \textbf{262}, 263 (1985).

\bibitem{friedberg1987}
R.~Friedberg, T.~D.~Lee, and Y.~Pang,
\newblock ``Mini-soliton stars,''
\newblock Phys.\ Rev.\ D \textbf{35}, 3658 (1987).

\bibitem{evstafyeva2026}
T.~Evstafyeva, N.~Siemonsen, and W.~E.~East,
\newblock ``Assessing the stability of ultracompact spinning boson stars with nonlinear evolutions,''
\newblock Phys.\ Rev.\ D \textbf{113}, 044024 (2026).

\bibitem{kaup1968}
D.~J.~Kaup,
\newblock ``Klein--Gordon geon,''
\newblock Phys.\ Rev.\ \textbf{172}, 1331 (1968).

\bibitem{ruffini1969}
R.~Ruffini and S.~Bonazzola,
\newblock ``Systems of self-gravitating particles in general relativity and the concept of an equation of state,''
\newblock Phys.\ Rev.\ \textbf{187}, 1767 (1969).

\bibitem{liebling2023}
S.~L.~Liebling and C.~Palenzuela,
\newblock ``Dynamical boson stars,''
\newblock Living Rev.\ Relativ.\ \textbf{26}, 1 (2023).

\bibitem{alic2012}
D.~Alic, C.~Bona-Casas, C.~Bona, L.~Rezzolla, and C.~Palenzuela,
\newblock ``Conformal and covariant formulation of the Z4 system with constraint-violating modes,''
\newblock Phys.\ Rev.\ D \textbf{85}, 064040 (2012).

\bibitem{grchombo}
T.~Andrade \emph{et al.},
\newblock ``GRChombo: an adaptable numerical relativity code for fundamental physics,''
\newblock J.\ Open Source Softw.\ \textbf{6}, 3703 (2021).

\bibitem{grteclyn}
GRTL Collaboration,
\newblock \texttt{GRTeclyn},
\newblock \url{https://github.com/GRTLCollaboration/GRTeclyn}.

\bibitem{grtresna}
J.~C.~Aurrekoetxea, S.~E.~Brady, \emph{et al.},
\newblock ``GRTresna: an open-source code to solve the initial data constraints in numerical relativity,''
\newblock J.\ Open Source Softw.\ \textbf{11}, 9057 (2026).

\bibitem{amrex}
W.~Zhang \emph{et al.},
\newblock ``AMReX: a framework for block-structured adaptive mesh refinement,''
\newblock J.\ Open Source Softw.\ \textbf{4}, 1370 (2019).

\bibitem{schoen_yau1979}
R.~Schoen and S.-T.~Yau,
\newblock ``On the proof of the positive mass conjecture in general relativity,''
\newblock Commun.\ Math.\ Phys.\ \textbf{65}, 45 (1979).

\bibitem{witten1981}
E.~Witten,
\newblock ``A new proof of the positive energy theorem,''
\newblock Commun.\ Math.\ Phys.\ \textbf{80}, 381 (1981).

\bibitem{mbarek2014}
S.~Mbarek and M.~B.~Paranjape,
\newblock ``Negative mass bubbles in de Sitter spacetime,''
\newblock Phys.\ Rev.\ D \textbf{90}, 101502(R) (2014).

\bibitem{gonzalez2009}
J.~A.~Gonz\'alez, F.~S.~Guzm\'an, and O.~Sarbach,
\newblock ``Instability of wormholes supported by a ghost scalar field. II. Nonlinear evolution,''
\newblock Class.\ Quantum Grav.\ \textbf{26}, 015011 (2009); see also \textbf{26}, 015010 (2009).

\bibitem{dzhunushaliev2014}
V.~Dzhunushaliev, V.~Folomeev, C.~Hoffmann, B.~Kleihaus, and J.~Kunz,
\newblock ``Boson stars with nontrivial topology,''
\newblock Phys.\ Rev.\ D \textbf{90}, 124038 (2014).

\bibitem{carroll2003}
S.~M.~Carroll, M.~Hoffman, and M.~Trodden,
\newblock ``Can the dark energy equation-of-state parameter $w$ be less than $-1$?,''
\newblock Phys.\ Rev.\ D \textbf{68}, 023509 (2003).

\bibitem{cline2004}
J.~M.~Cline, S.~Jeon, and G.~D.~Moore,
\newblock ``The phantom menaced: constraints on low-energy effective ghosts,''
\newblock Phys.\ Rev.\ D \textbf{70}, 043543 (2004).

\bibitem{helfer2022}
T.~Helfer, U.~Sperhake, R.~Croft, M.~Radia, B.-X.~Ge, and E.~A.~Lim,
\newblock ``Malaise and remedy of binary boson-star initial data,''
\newblock Class.\ Quantum Grav.\ \textbf{39}, 074001 (2022).

\bibitem{cttk}
J.~C.~Aurrekoetxea, K.~Clough, and E.~A.~Lim,
\newblock ``CTTK: a new method to solve the initial data constraints in numerical relativity,''
\newblock Class.\ Quantum Grav.\ \textbf{40}, 075003 (2023).

\bibitem{bona1995}
C.~Bona, J.~Mass\'o, E.~Seidel, and J.~Stela,
\newblock ``New formalism for numerical relativity,''
\newblock Phys.\ Rev.\ Lett.\ \textbf{75}, 600 (1995).

\bibitem{campanelli2006}
M.~Campanelli, C.~O.~Lousto, P.~Marronetti, and Y.~Zlochower,
\newblock ``Accurate evolutions of orbiting black-hole binaries without excision,''
\newblock Phys.\ Rev.\ Lett.\ \textbf{96}, 111101 (2006).

\bibitem{baker2006}
J.~G.~Baker, J.~Centrella, D.-I.~Choi, M.~Koppitz, and J.~van~Meter,
\newblock ``Gravitational-wave extraction from an inspiraling configuration of merging black holes,''
\newblock Phys.\ Rev.\ Lett.\ \textbf{96}, 111102 (2006).

\bibitem{trivedi_loeb2026}
O.~Trivedi and A.~Loeb,
\newblock ``Unique gravitational-wave signals from negative-mass binaries,''
\newblock Phys.\ Dark Univ.\ \textbf{53}, 102409 (2026).

\bibitem{battye_sutcliffe2000}
R.~A.~Battye and P.~M.~Sutcliffe,
\newblock ``Q-ball dynamics,''
\newblock Nucl.\ Phys.\ B \textbf{590}, 329 (2000).

\bibitem{palenzuela2008}
C.~Palenzuela, L.~Lehner, and S.~L.~Liebling,
\newblock ``Orbital dynamics of binary boson star systems,''
\newblock Phys.\ Rev.\ D \textbf{77}, 044036 (2008).

\bibitem{manfredi2018}
G.~Manfredi, J.-L.~Rouet, B.~Miller, and G.~Chardin,
\newblock ``Cosmological structure formation with negative mass,''
\newblock Phys.\ Rev.\ D \textbf{98}, 023514 (2018).

\bibitem{manfredi2026}
G.~Manfredi, J.-L.~Rouet, and B.~Miller,
\newblock ``The Bondi universe: can negative mass drive the cosmological expansion?,''
\newblock Phys.\ Dark Univ., 102493 (2026).

\bibitem{repo}
N.~M.~Shirokov,
\newblock \texttt{GRTeclyn} research repository and data release,
\newblock \url{https://github.com/Nikchik-coder/GRTeclyn}.

\bibitem{shirokov2026}
N.~M.~Shirokov,
\newblock ``Wormhole dynamics: nonlinear collapse and gravitational-wave emission,''
\newblock arXiv:2604.00071 [gr-qc] (2026).

\bibitem{shirokov_merger2026}
N.~M.~Shirokov and I.~Nachevsky,
\newblock ``Binary Ellis--Bronnikov wormholes in $3{+}1$ numerical relativity: mergers, fly-bys and gravitational waves,''
\newblock arXiv:2610.13206 [gr-qc] (2026).

\bibitem{shirokov_spin2026}
N.~M.~Shirokov and I.~Nachevsky,
\newblock ``Rotation slows the instability of Ellis--Bronnikov wormholes,''
\newblock arXiv:2610.29941 [gr-qc] (2026).

\end{thebibliography}
\end{document}